# Electronic structure and physical properties of ¹³C carbon composite

ABSTRACT: This review is devoted to the application of graphite and graphite composites in science and technology. Structure and electrical properties, as so technological aspects of producing of high-strength artificial graphite and dynamics of its destruction are considered. These type of graphite are traditionally used in the nuclear industry. Author was focused on the properties of graphite composites based on carbon isotope ¹³C. Generally, the review relies on the original results and concentrates on actual problems of application and testing of graphite materials in modern nuclear physics and science and its technology applications.

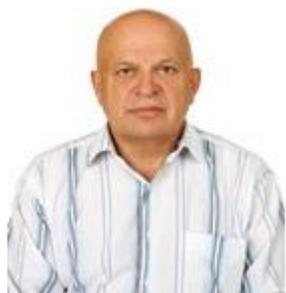

Author: *Evgenij I. Zhmurikov*
Address: *68600 Pietarsaari, Finland*
E-mail: *evg.zhmurikov@gmail.com*

## 1. Introduction

In order to explore ever-more exotic regions of the nuclear chart, towards the limits of stability of nuclei, European nuclear physicists have built several large-scale facilities in various countries of the European Union. Today they are collaborating in planning of a new radioactive ion beam (RIB) facility which will permit them to investigate hitherto unreachable parts of the nuclear chart. This European ISOL (isotope-separation-on-line) facility is called EURISOL [1].

At the present moment experiments with RIB of the first generation have yielded important results. But the first generation RIB facilities are often limited by the low intensity of the beams. This is why there have been discussed and developed quite a number of new projects with a new generation RIB (KEK, Tsukuba; ARENAS REX ISOLDE, CERN; ARGONNE; HRIBF, Oak Ridge). The SPES project at LNL (Italy) also belongs here [2, 3].

The proposed project is aimed at an R&D study of a conversion element in the framework of the SPES program at LNL. LNL has proposed a double-acceleration scheme of producing an ISOL type RIB. A primary proton beam accelerated at a superconducing RFQ linac is directed to a special neutron target and produces an intense ($3 \times 10^{14}$ cm$^{-2} \times $s$^{-1}$) flux of fast neutrons. The parameters of the primary beam are: the energy is up 100 MeV, the average power is up to 300 kW, the diameter is 1 cm. Thus obtained neutron flux then comes to a hot thick target of a $^{238}$U compound. The vapors of the radionuclides get ionized, extracted from the target at an energy of 20-60 keV and then being separated by isotopes get into the experimental zone for low-energy experiments or further accelerated to an energy of up to 1-5 MeV/nucleon. The conversion element consisting of a neutron target and a source of radioactive ions plays a the most important role in this scheme [4-7].

Version converter with a target made of a carbon composite with a high content of the isotope $^{13}$C was developed for the proton primary beam. The threshold of the $^{13}$C nuclear reaction $^{13}$C (p, n) $^{14}$N with yield neutrons is of the 3.24 MeV, while reaction threshold $^{12}$C (p, n) $^{13}$N reaction for usual carbon is significantly above and equal to 20.1 MeV [8, p .895]. It is supposed, the neutron yield from a carbon target that made of pure isotope $^{13}$C, can be considerably higher if energy of the proton beam is smaller than 20 MeV. However, it is turned out quite quickly that at higher energies this difference is not so noticeable [9]. Nevertheless, in principle, such a cooled by radiation target can provide in 3-10 times more high neutron yield than natural carbon $^{12}$C target at low energies of the proton beam. An argument in favor of such a target is that the using of the deutron beam, as well as an increase of its power, it will be resulting into a significant increase of the cost and complexity of the project EURISOL in whole.

Moreover, the $^{13}$C isotope composite finds out its application in the resonance gamma spectroscopy [10, 11]. This method is based on that the nuclei of many chemical elements have the property of resonance absorption. In particular, the nucleus of the nitrogen atoms can resonantly absorb gamma quanta with energy of 9.1724 MeV, and the width of the absorption peak is of 125eV only. Method of nitrogen detection is found in comparison to the resonance and non-resonant absorption when a radiation comes through the matter. Nuclear reaction $^{13}$C (p, $\gamma$) $^{14}$N is used to generate a resonance $\gamma$-quanta. In this case accelerated up to 1.75 MeV proton beam bombards the graphite target with a high content of $^{13}$C carbon isotope, therefore it is formed an excited nucleus $^{14}$N that emits gamma quanta with energy of 9.17 MeV. In this case, the necessary energy for resonance spectrometry has gamma-quanta with emission angle $80,7 \pm 0,1^0$ to the axis of the proton beam [11].

This review summarizes the results of studies of an electronic structure and properties of $^{13}$C carbon composite, because it can allow to predict its lifetime and its possibilities as a construction material. Samples of this composite with a high content of the $^{13}$C isotope were made in "NIIgraphit"[12] of powder that was obtained by chemical vapor deposition. Properties of the graphitization and a technological scheme of the graphite preparation were described earlier [13, 14].

## 2. X-ray and high-resolution microscopy of the $^{13}$C carbon composite

X-Ray measurements and high-resolution microscopy measurements were carried out and described by leading of prof. *S.V. Tsybulya* in the Boreskov Institute of Catalysis SB RAS and published earlier in [15]. It was used URD-6 with monochromatic Cu$K\alpha$-radiation. The registration of diffractograms was performed in steps mode with step 0.05°, the accumulation time is 10 seconds and the angle range 2$\theta$ was from 10 to 100°.

The XRD profile of the $^{13}$C carbon composite is shown on fig. 1. It is clear visible *00l* and *hk0* reflection (line 2) of graphite planes, the *hk0* reflexes have an asymmetrical shape with a large blurring towards larger angles than in the case of high-ordered polycrystalline graphite (line 1). This diffraction pattern complies to the turbostratic graphite structure in which there are not ordered graphene layers along the *c* - crystallographic direction [16]. The diffractogram of $^{13}$C carbon isotope really shows (line 3) really only one well-defined broad diffraction peak, that complies to the *002* reflex of the graphite structure. Enlarged image of the diffraction pattern allows us to identify a broad peak *100*, located at 2θ =44° (fig. 2). The size of the coherent scattering region CSR (Å) for the structural components of the $^{13}C$ powder isotope is shown in table 1.

Table 1. *The dimensions of the coherent scattering region (Å) for the structural components of the initial $^{13}C$ powder*

| Structural components | $d_{002}$ (Å) | 00l (Å) |
|---|---|---|
| $G_1$ | 3.43 – 3.44 | 40 |
| $G_0$ | 3.6 – 3.65 | 20 |

A more detailed structure of the *002* reflex of the $^{13}C$ powder is shown in the inset to fig. 2. The presence of an inflection indicates that the powder is composed of two structural components, which differ in size. Indeed, the *002* peak is a superposition of two components, relating to the graphite particles with CSR of 20Å and 40Å and different interplanar distance $d_{002}$ (table 1). Similar values of integral component of intensities indicate to equal proportion of these particles in the sample. Low intensity of *100* peak does not allow to attribute it to one or another of the structural component.

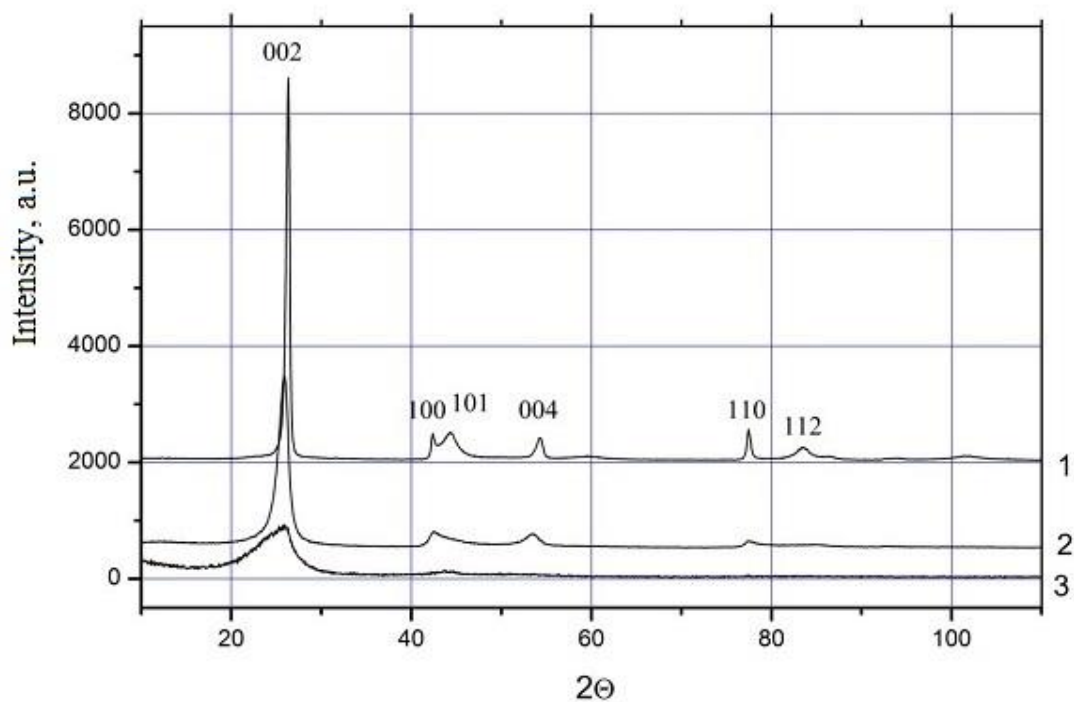

**Fig. 1**. *The X-Ray phase diagram: 1) MPG-6; 2) the initial tablet of $^{13}C$ carbon composite 3) initial powder of $^{13}C$ pure isotope* [17].

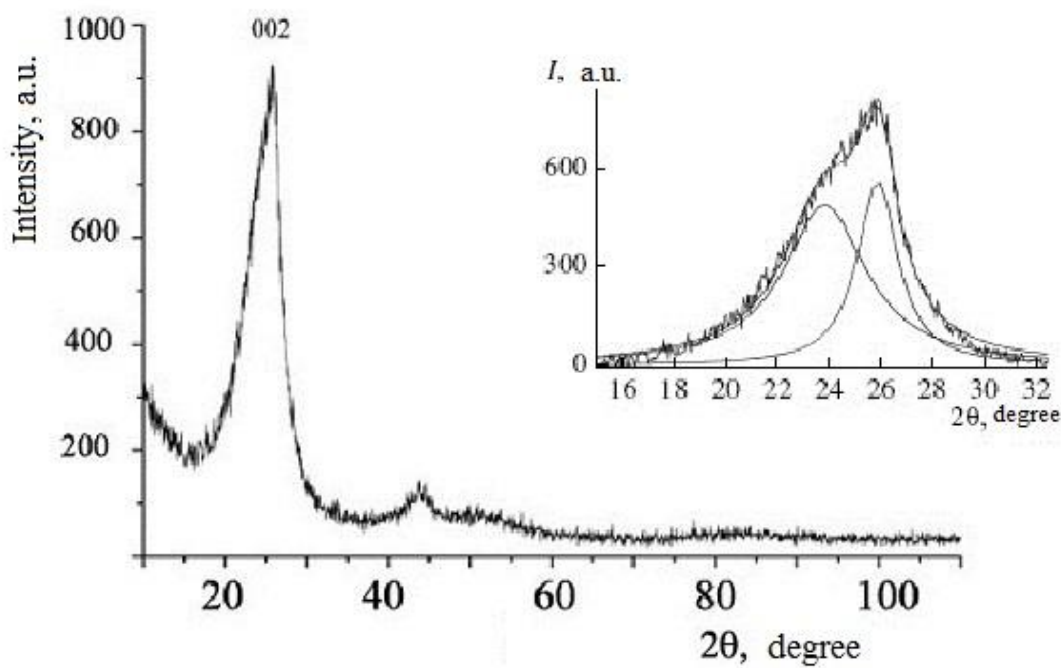

**Fig**. 2 *The X-Ray phase diagram of the initial powder of $^{13}C$ pure isotope. The shape of the diffraction 002 peak (inset at top) indicates the presence in the sample of two structural components with different dispersion and, probably, with different interplanar spacing $d_{002}$* [17].

High-resolution transmission electron microscopy (HRTEM) is shown that the sample of $^{13}C$ carbon composite of density less than 0.8 g/cm$^3$ consists of particle aggregates with size up to 1000 nm (fig. 3a). However, the morphology of particles that make up the aggregate is really different from carbon composite of MPG class, because each thin plate is arranged so, that looks like a sheet of paper, that was crumpled in the middle, and then a little straightened [15].

Structure of the plate is not a monocrystalline, and complies a polycrystalline state: a set of randomly oriented interconnected blocks, giving the ring microdiffraction that is shown on the inset to fig. 3 a. It is interesting that curving edges of such plates are similar to carbon fiber that can be clearly seen in the micrograph of larger scale (fig. 3,b). Constructional composites are based on $^{13}C$ isotope with high density ($\rho \sim 1.55$ g/cm$^3$), and isotope content of from 50% up to 75% are composed of three types of carbon particles.

The main mass of the samples are the crumpled and broken graphite plate with thickness from of 1 to 50 nm, which have a tendency to agglomerate. Moreover, the sample contains graphite globules, which are generally well faceted and have the sizes from 50 to 150 nm; some globules are partially destroyed. Every facet of the globule is a graphite plate of thickness about 15–20 nm.

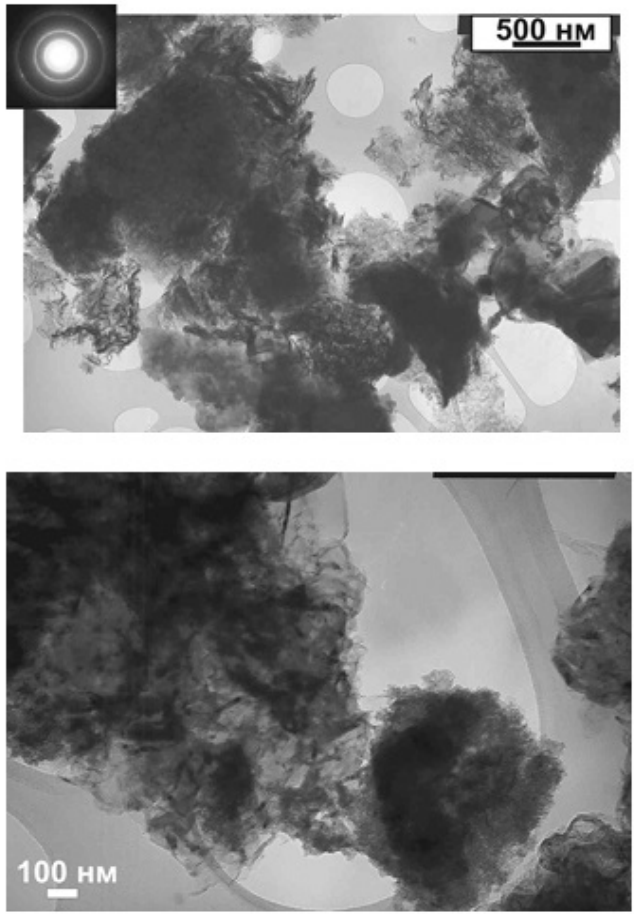

**Fig**. **3**. *The micrographs and microdiffraction of $^{13}C$ carbon composite of density less than 0.8 g/cm³* [15].

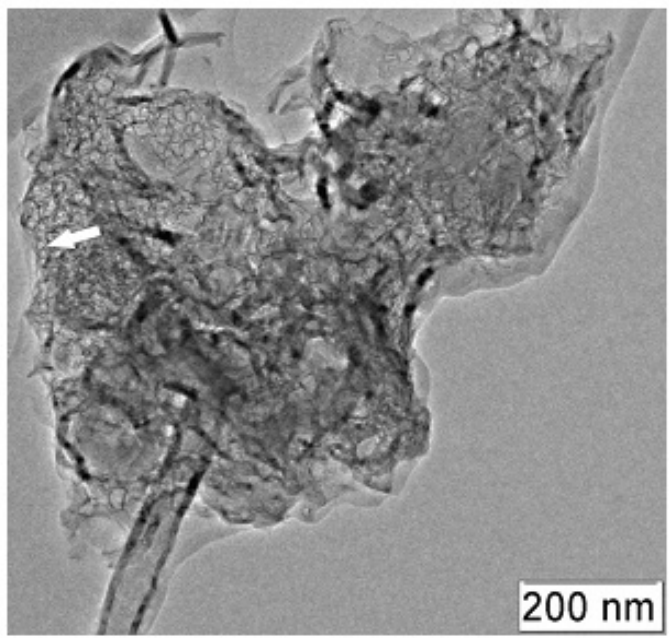

**Fig**. **4** *Image of $^{13}C$ carbon composite structure with morphological type of "moire". The arrow indicates a point from which a high-resolution a snapshot was obtained* [17].

Fig. 4 shows a high-resolution image of formations, that called "*moire*". This form of carbon is composed by the crumpled graphene layer of thickness about 2 nm, that are interspersed of thick graphite plates at 10 nm and a length of 0.1 mkm about. An enlarged image of such formations is shown in fig. 5.

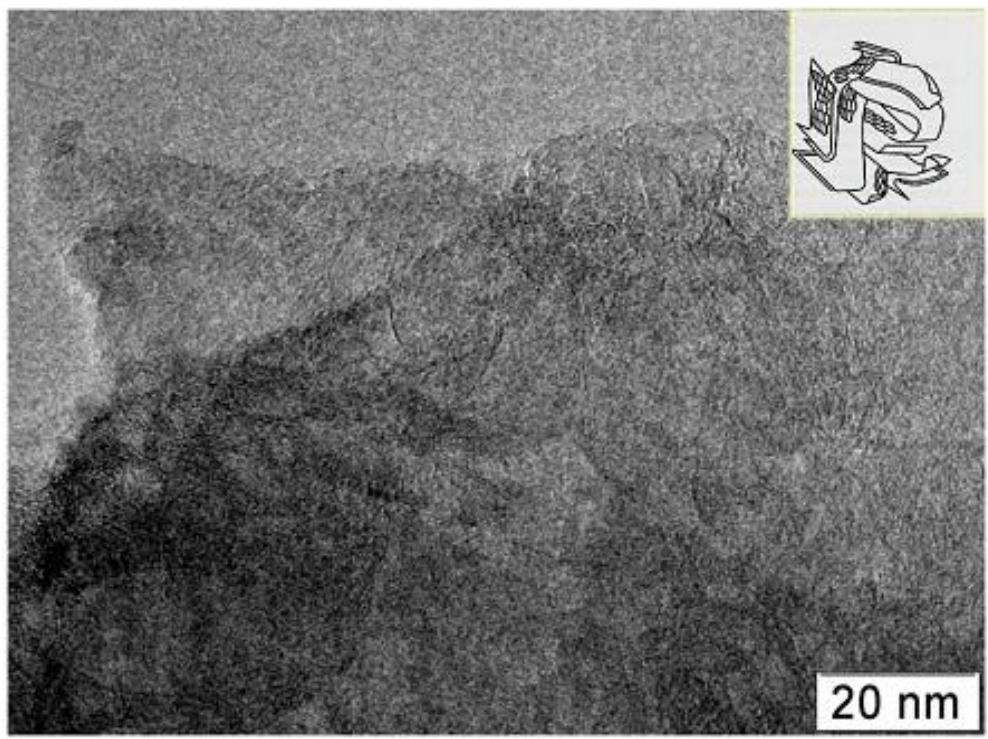

**Fig. 5** *The "moire" morphological structure of the $^{13}C$ carbon composite is shown on the high-resolution image. The material is composed of randomly arranged curved and fractured sheets of graphite, that in turn comprised of 10-20 graphene layers. The turbostrate model of graphite structure can be represented as a "lump of crumpled paper".* [17]

3. **The electronic structure investigations of $^{13}C$ carbon composite by X-ray fluorescence spectroscopy method and quantum-chemical simulation**.

This measurements were carried out and described by leading of prof. *A.V.Okotrub* in the Nikolaev Institute of Inorganic Chemistry of the SB RAS and published earlier in [17]. X-ray fluorescence *CKα* spectra of polycrystalline graphite, $^{13}C$ powder and $^{13}C$ composite have been measured on the laboratory spectrometer. As the crystal-analyzer was used a biphthalate ammonium monocrystal; it was applied the mathematical procedure describing in [18] to account of the nonlinear crystal reflection efficiency. Samples were deposited onto a copper substrate and cooled in a vacuum chamber of X-ray tube up to liquid nitrogen temperature. Operating mode of X-ray tube was U = 4 kV, J = 0.8 A. Register of X-rays was carried out of the gas proportional counter that was filled of methane at a pressure of about 0.1 atm. The resolution of the spectrometer is 0.4 eV. Energy of X-ray line was defined with an accuracy of ± 0.15 eV.

X-ray fluorescence emission spectrum is the result of a vacancies filling that were pre-created in the atomic energy levels of the valence electrons. The X-ray transition in the carbon compound due to the dipole selection rules is carried between *2p* and *1s-* orbitals of the carbon atom. Consequently, *CKα*-spectrum has an information about of density of C*2p* states in the valence band of a compound. *CKα*-spectra of polycrystalline graphite, $^{13}$C powder and $^{13}$C composite are shown on (fig. 6a). Analysis *CKα* spectrum of graphite was based on the interpretation of [19]. The spectra are three basic features that can be marked by the letters **A**, **B**, **C**. Intensive maximum of **C** (*E* = 276.5eV) complies of *σ*-electron of graphite, high-energy maximum of **A** (*E* = 282 eV) refers to the *p*-system, and **B** maximum is formed by both types of electrons.

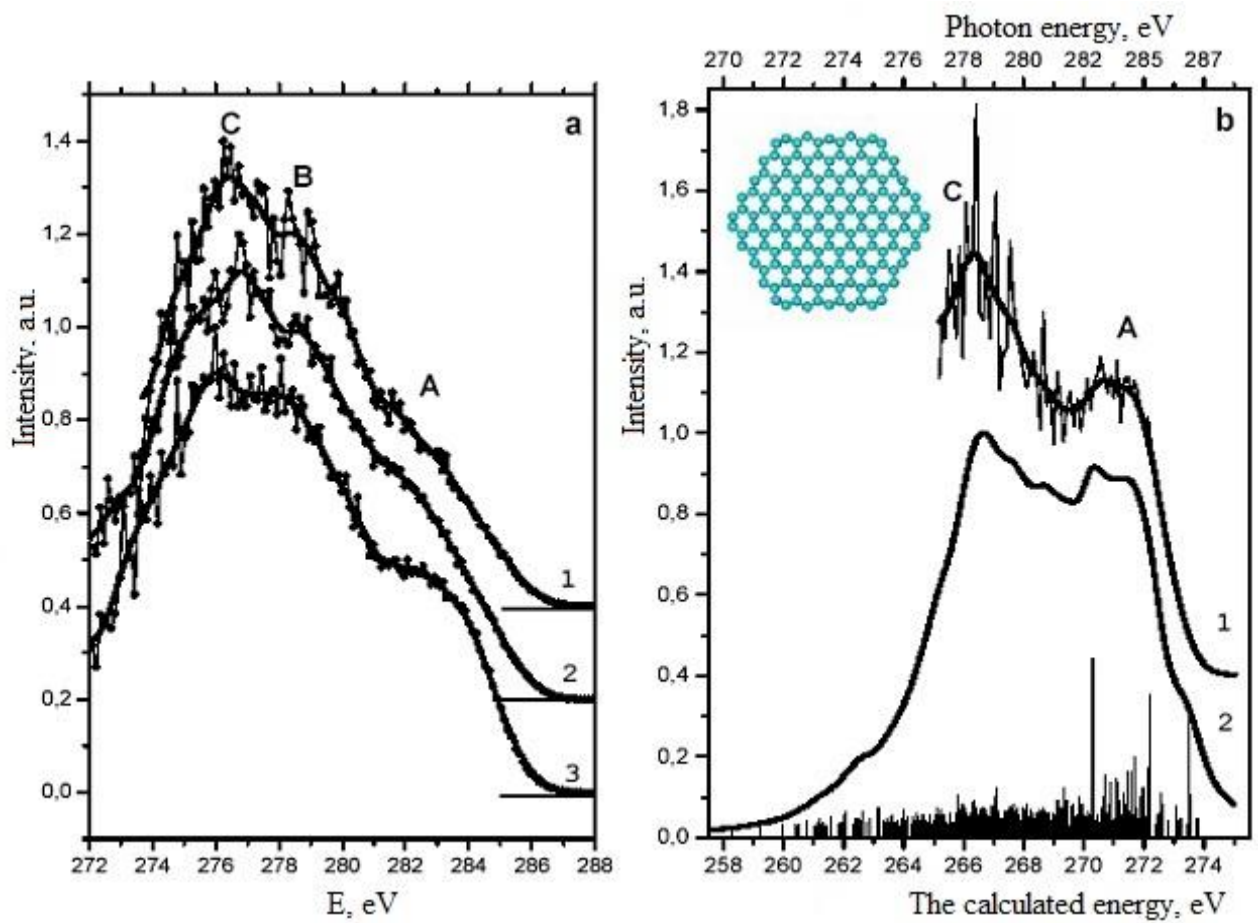

**Fig**. **6**. **a**). *Comparison of CKα spectra that were measured for the natural graphite (1); for the $^{13}$C powder (2) and for $^{13}$C composite (3). b) Comparison of the CKα spectrum, that was obtained by subtracting of 60% of the intensity of the natural graphite spectrum from the CKα spectrum of the $^{13}$C powder (1); the theoretical CKα spectrum of the $C_{150}$ graphene fragment (2); the structure of this fragment is represented in the upper left corner of the picture.*

*CKα*-spectrum of $^{13}$C composite is quite similar to spectrum of natural graphite in the position and the relative intensity of the main features (fig. 6, line 2), that agrees with the data of a X-ray diffraction. Increasing of the relative intensity maxima **B** and **A** in comparison with that one of the ordinary graphite is observed in the spectrum of the $^{13}$C powder (fig. 6,a, line 3). Increasing density of high-energy states in spectrum of the $^{13}$C powder can be due to the presence of large number of defects, that disturb of the homogeneity of the carbon hexagonal grid [20, 21].

XRD analysis of the $^{13}$C powder (fig. 1) was revealed the presence in a sample of the structural components $G_0$ and $G_1$ with the size of about 20Å and 40Å. However, these size can be typical for the graphene fragments that make up these particles. It was proposed that *CKα* spectrum of large particles (~ 40 Å) is quite close to the spectrum of natural graphite. From *CKα* spectrum of $^{13}$C powder was subtracted ~ 60% of the intensity *CKα* spectrum of natural graphite to separate electron state of of small particles (~ 20 Å). The normalized result of a subtraction is shown on (fig. 6, b).

The features of the resulting spectrum in this approach are more clearly identify in comparison with the initial spectrum of natural graphite, $^{13}$C powder and $^{13}$C composite. This normalized spectrum shows an increase of the intensity of the **A** line and the maximum of this line shifts to a high-energy region.

It was calculated theoretical spectrum C*Kα* graphene structure, consisting of 150 carbon atoms to confirm that the resulting spectrum relates to small-sized graphite particles. The size of this fragment was equal about ~ 20 Å. The geometry of $C_{150}$ fragment has been optimized in the DFT with using of three-parameter hybrid functional Becke [22] and correlation functional Lee, Yang and Parr [23] (B3LYP method) in frame of the package of quantum chemical programs *jaguar* [23]. Atomic orbitals are described by 6-31G** basis set. Energy of X-ray transition was defined as the an energy difference between the one-electron *i* and *j* levels, if *j* is an internal level, and the *i* is level that belongs to the valence band of the molecule. The line intensity, that complies to the X-ray transition, calculated from the formula:

$$I_{ij} = \sum_A \sum_n \sum_m \left| C^A_{jm} C^A_{in} \right|^2 \qquad (1),$$

where **A** is carbon atom of molecule, $C^A{}_{jm}$ and $C^A{}_{in}$ are the coefficients, with which *1s*-AO and *2p*-AO are included into the *c*alculation of *i*-th and *j*-th MO. Each line in the spectrum has been expanded of Lorentz 0,6eV-lines and normalized to the maximum value. Comparison was made between the theoretical spectrum, calculated for atoms fragment $C_{150}$ (fig.6, b) and *CKα* spectrum, that was obtained by subtracting the spectrum of natural graphite from the spectrum of $^{13}$C powder. Both theoretical and experimental spectra are shown two peaks **C** and **A**, the distance between them and the relative intensity of the spectra are similar for both.

The high intensity of **A**-peak is due to the contribution of the electrons of broken bonds of the two-coordinated atoms, that a fragment border are composed. The highest electron localization is typical for the atoms, that make up a zigzag edge [25]. The intensity of the **A**-peak in the calculated spectrum a little overstated in comparison with the experimental one. This fact could indicate a little larger size of graphite particles of $^{13}$C powder than the size of a calculated cluster $C_{150}$.

### 4. Electrophysical measurement

The electrophysical measurement were carried out and described by leading of prof. *A.I. Romanenko* in the Nikolaev Institute of Inorganic Chemistry of the SB RAS and published earlier in [17]. Temperature dependence of the electrical conductivity of the $^{13}$C composites samples were measured by the four-probe method at a constant current.

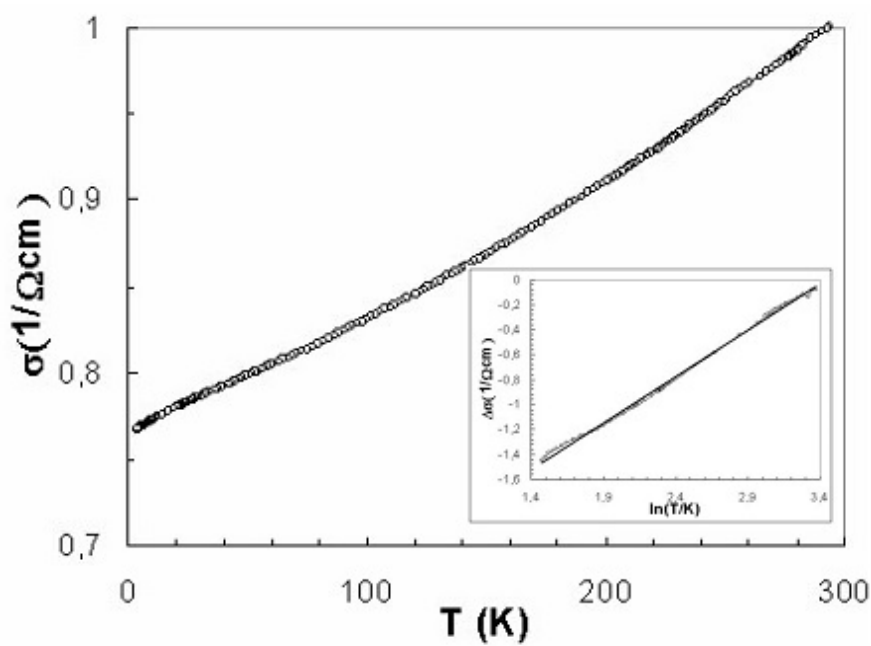

**Fig. 7** *The temperature dependence of a conductivity of the $^{13}$C composite with high density ρ ~ 1.55 g/cm³. Inset: anomalous part of low temperature dependence of the electrical conductivity of $^{13}$C carbon composite in the coordinates Δσ(T)– lnT* [17].

The sample for measurement was a block with the size of 1 × 10 × 0.5 mm³, cut out from the initial carbon material. The electrical contacts were plotted by silver paste *G3692 Acheson Silver DAG 1415 mit Pinsel* (Germany) and have a resistance of 1 Ohm about. Measurements were carried out in helium/air atmosphere in the temperature range of 4.2–300 K.

The temperature dependence of conductivity of the $^{13}$C dense constructional composite with $\rho \sim 1{,}55$ g/cm$^3$ is close to linear (fig. 7). It can be explained by the dominance of a three-dimensional quantum corrections in the entire temperature range due to the weak localization of carriers [26, 27]. The magneto-resistance measurements of $^{13}$C composite at helium temperatures also indicates to the determining role of quantum corrections [17]. Magneto-resistance is negative over the entire temperature range from 0 up to ± 1.15 Tl and it has a logarithmic asymptotic, which is without doubt associated with the suppression of a interference quantum correction of a magnetic field.

The so-called anomalous part in the low temperature part of the conductivity temperature dependence can be allocated (fig.7, inset), using the procedure that was set out in [28]. In this case:

$$\Delta\sigma(T) = \sigma(T)_{experimental} - \sigma(T)_{extrapolation} \qquad (2).$$

The anomalous part, that was obtained by subtracting a regular part from of the experimental data is quite well approximated by a direct in the coordinates $\Delta\sigma(T) - \ln T$. It is also indicate of the fact, that the 2D weak-localization quantum corrections are dominated at low temperatures. A possible reason for the deviation of the conductivity temperature dependence from the logarithmic with an increasing of temperature can be non-zero probability of a carriers transition between the entangled, broken and high-defected graphene layers (fig.4).

The quantum corrections of this kind of electron transition are analyzed in [29] in the case of oxide superconductors with of a perefskito-like layered structure. In accordance with [29] the influence of the carrier transitions between the layers can be qualitatively explained as follows. The total value of the interference correction is proportional to the probability of returning to the the start point for the time that is shorter than dephasing time $\tau_\varphi$.

It was supposed that the structure represents two parallel layers. If the transitions between them are not, each layer gives the correction to the conductivity according to the formula (3) in the form:

$$\Delta\sigma/\sigma \sim -e^2/\hbar \ln(L_\varphi/l) \qquad (3).$$

In this case, the diffusion length is connected to the relaxation time $\tau_\varphi$ of its wave function relation:

$$L_\varphi = (D\tau_\varphi)^{1/2} \qquad (4).$$

The total correction to the conductivity of a two-layer structure is respectively twice as large. The electron can go in an adjacent layer with the coordinate *z* instead of returning to the starting point, if it is appeared that the transition time between layers $\tau_{ij}$ is comparable to the time of failure of its wave function $\tau_\varphi$. It is clear that such type of a trajectory is not longer contribute to the interference, and the quantum correction value will be less than $-2e^2/\hbar \ln(L_\varphi/l)$.

Furthermore, it will be change the shape of the conductivity temperature dependence, of course. The three-dimensional Fermi's quasi-surface is formed in the case, that the time of electron transition between layers $\tau_{ij}$ is less than energy relaxation time $\tau_\varepsilon$, [29]. Then the quantum corrections for such type of a structure should be considered as well as for three-dimensional anisotropic conductor.

## 5. Thermophysical measurements

Thermophysical measurements were carried out by leading of prof. *S.V. Stankus* in the Kutateladze's Institute of Thermophysics SB RAS and were interpreted by *E.I. Zhmurikov* earlier in [30]. This study was performed of the thermal laser flash method on the automated experimental setup LFA-427 of company Netzsch [31] and it was described earlier in detail in [30, 32]. The main advantages of LFA-427 are the wide temperature range, available for measurement (25 ...2000C); the opportunity to explore different classes of solids; the small size of the sample (thickness – 0.1 ...6 mm, diameter – 6 ...12 mm); possibility to operate in a vacuum ($10^{-5}$ Torr), oxidative and protective (Ar, He) atmospheres; a wide range of the thermal diffusivity measurements (0.01 ... 10 cm$^2$/s); high (2 ... 5%) accuracy and efficiency of measurement; availability of automated control systems and data processing. The estimated error of thermal conductivity measurements is 3 – 6% about.

The results of the temperature dependence of a thermal conductivity for different type of graphite are shown on fig.8. It is clearly visible, that the temperature dependence of a thermal conductivity of MPG-6 or SGL graphite are weakly nonlinear and the thermal conductivity of $^{13}$C composite practically do not varies with temperature.

By now it is well known, that the phonon-phonon interaction and U-processes almost completely are responsible of the temperature dependence of a graphite thermal conductivity at high temperatures, as well as the scattering of phonons by the boundaries of the crystallites, inheterogeneous structure and lattice defects [33–36].

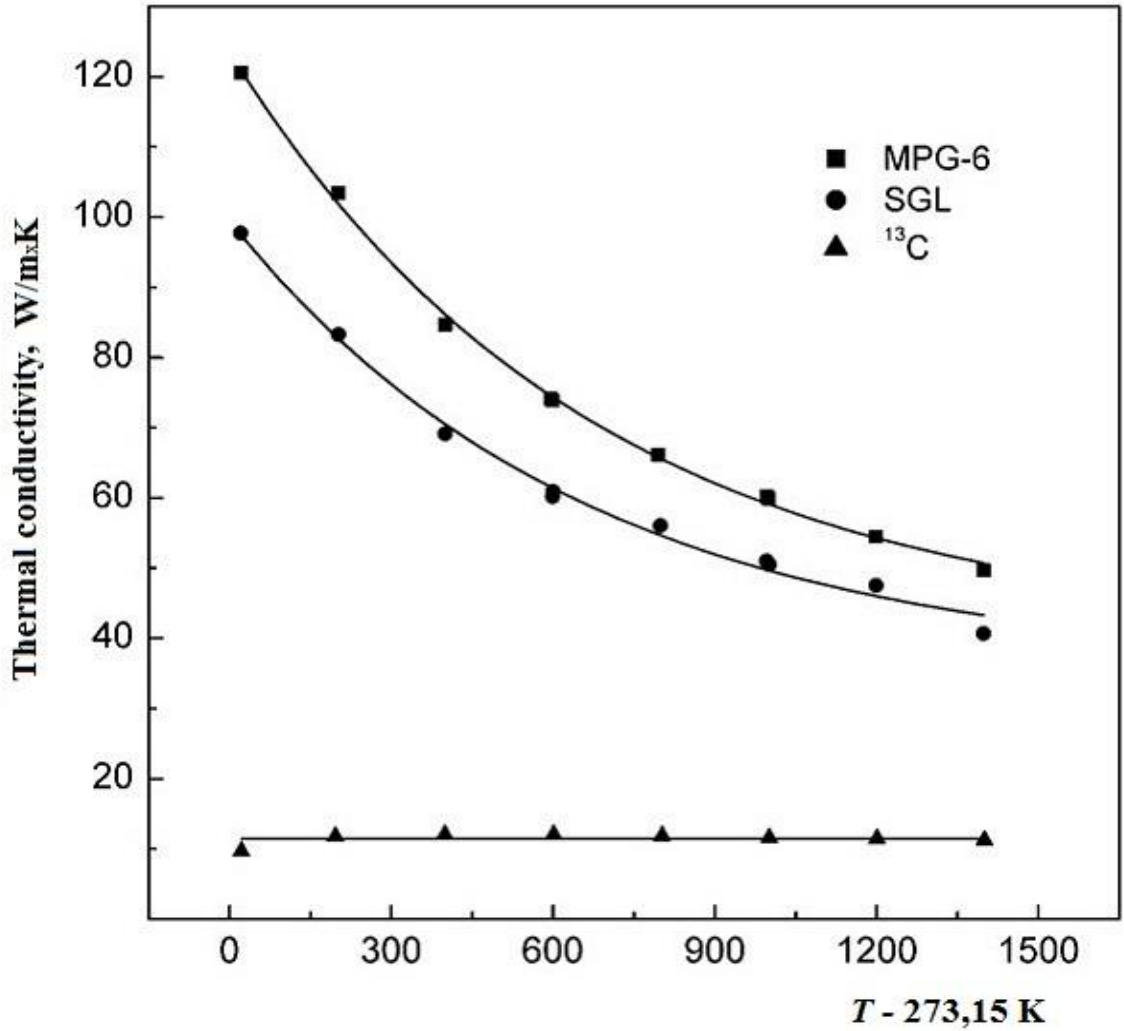

**Fig**.8. *Temperature dependence of the thermal conductivity of graphite. It was calculated from measurements of the thermal diffusivity of graphite composites* [32].

A study of the crystal structure and the electronic properties of both natural and the $^{13}$C graphite composites has been made earlier by high-resolution transmission electron microscopy (HRTEM), X-ray diffraction, X-ray fluorescence spectroscopy, Raman scattering, etc. [15, 17, 37-39]. The sharp distinctions were revealed between the comparably perfect polycrystalline structure of the MPG graphite and the small-grained turbostratic structure of the $^{13}$C carbon composite.

In particular, in [37], from the data of X-ray diffraction analysis and high-resolution electron microscopy it was concluded that the MPG graphite consists of large units (over 1000 nm). These units were formed by fine-faceted irregular polyhedral plates and the average size of the coherent scattering region (CSR) was about 1500Å.

The phenol formaldehyde resol resin was used as a binder in the case of the $^{13}$C carbon composite and the relatively low temperature of the final graphitization stage (⩽ 2200$^0$C) to produce the structure which is morphologically similar to that of glassy carbon [13]. The macromolecular, polyhedral structure of glassy carbon was analyzed in detail and described early in [40]. The transmission electron microscopy shows [15] that in the $^{13}$C composites, the coherent scattering domain (CSD) size is about ~ 100Å only. According to the electrical measurements, the coherence length at helium temperature is also about 150Å. Thus, the average size of crystallites can be estimated as 100-150 Å.

So, it seems, that in this case as well as in the case of glassy carbon, a sharp decrease in thermal conductivity with temperature is associated with the fact that the phonons mean free path (MFP) is limited of the microcrystallite size and is not dependent on temperature (fig. 8).

### 6. Features of the graphite phonon spectra

The phonon spectra features are discussed also in [30]. In the crystalline graphite, heat transfer is performed by phonons over the entire temperature range [34, 41-48]. The only exception is the low-temperature range up to about 10K, where the electron contribution to thermal conductivity in a transverse magnetic field can reach 40% [34]. In turn, the particular features of the graphite phonon spectra of graphite are entirely determined by the peculiarities of the graphene phonon spectrum. Graphene has a hexagonal structure with two carbon atoms in an elementary cell, which results in six phonon polarization branches in the dispersion spectrum. The phonon modes $LA$ and $TA$ comply with the longitudinal and transverse vibrations of carbon atoms in graphene planes. The out-of plane mode $ZA$ responds to vibrations of carbon atoms in the perpendicular direction to the $LA$ and $TA$ vibration modes. The acoustic $LA$ and $TA$ modes have a linear dispersion, and the sound speeds of these modes are ($LA$)=2.13×10$^6$ cm/s and ($TA$)=1.36×10$^6$ cm/s, respectively [41]. There is no consensus now about the dispersion of the so-called «bending» $ZA$ mode. In [43], it is assumed to be the quadratic character dispersion at $\Gamma$- point of the Brillouin zone. At the same time, the calculation results [45] shows in favor of the linear dispersion of the $ZA$ mode, that is characterized by the sound speed ($ZA$) = 0.16 ×10$^6$ cm/s [42].

The Debye heat capacity theory cannot be considered as rigorous by its basic assumption. In particular, it uses only one acoustic branch, limiting the approach of the continuum and assuming that there are no optical modes and three acoustic branches are equal. In addition, the Debye theory assumes only the linear dispersion, and thus, takes no account of the phonon-phonon interaction. From a physical point of view, this means that the elastic deformation waves propagate through a crystal without interacting with each other, and thermal energy is transferred by phonons with sound velocity.

One of the major simplifications of the Debye theory is the choice of a quadratic dependence for the spectral density $G(\omega)$, which can differ drastically from its true form [46-48] except for the region of very low frequencies. The *ab initio* calculation for phonon state density according to the GGA method (Generalized Gradient Approximation) is exemplified in [46].

The phonon states density finds out the best reflection in the spectra of the first- and second-order Raman scattering of crystal and pyrolytic graphite [48]. In this model, the *D*-band Raman spectrum uniquely is associated with the phonon density of states at *K*-point of the Brillouin zone and the $I(D)/I(G)$ ratio of band intensities are determined fully by the size of the crystallites. At the same time, one cannot exclude that the appearance of new bands in the Raman spectrum can be attributed to the formation of new carbon polymorphs, which may be assigned, e.g., to the presence of $sp^3$-bonds in clusters [48].

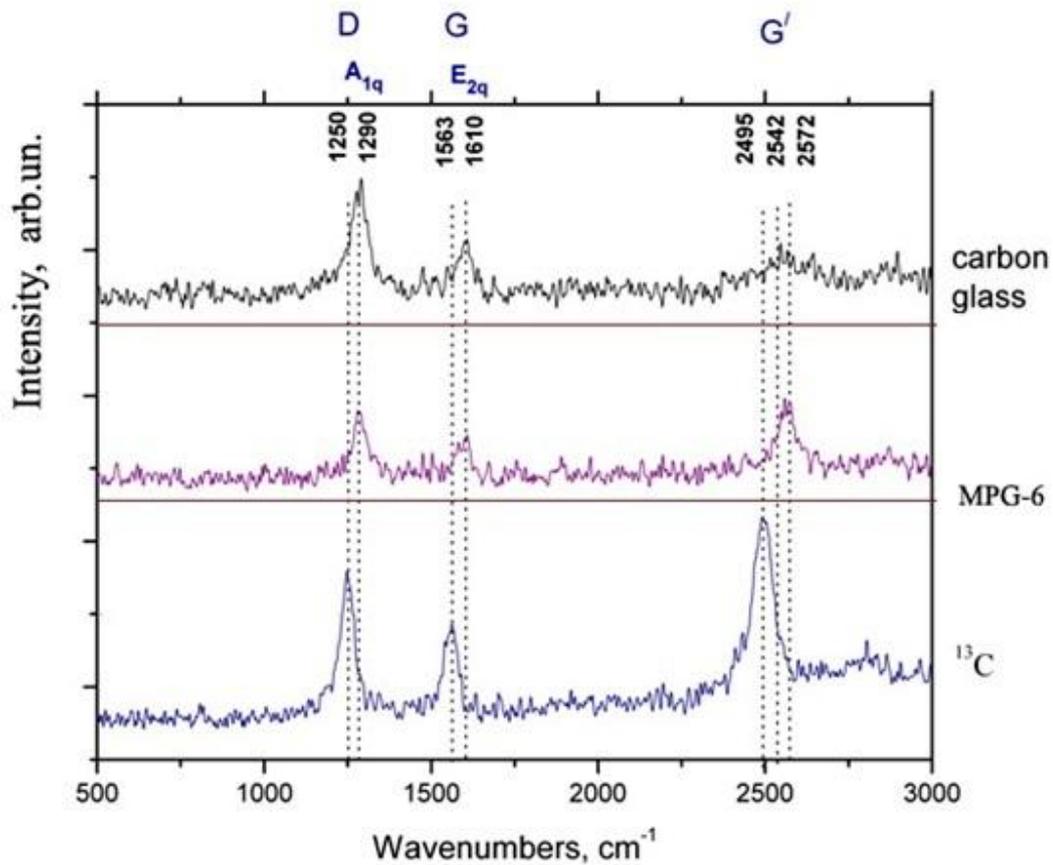

**Fig**. **9** *Raman scattering spectra of carbon materials from top to bottom: glassy carbon; MPG-6; $^{13}C$ carbon composite* [15]

The high-frequency *LO* and *TO* optical modes of graphene are degenerate at $\Gamma$-point of the Brillouin zone and belong to the $E_{2g}$ two-dimensional representation. These modes with a characteristic frequency of 1580 cm$^{-1}$ are used in graphite studies that was performed by Raman scattering method.

Interpretation of the Raman scattering spectra of carbonaceous materials is based on the analysis of the relative intensities of oscillation lines complies to the active Raman vibration type E$_{2g}$ (1580 cm$^{-1}$) with the symmetry of the unit cell D$_{6h4}$ for the perfect graphite crystal. The A$_{1g}$ oscillation (1355 cm$^{-1}$) is forbidden by selection rules for the ideal graphite, but appears in the Raman scattering spectra of the graphite crystals of limited size and amorphous carbon forms (1355 cm$^{-1}$) [49, 50].

Raman spectra were measured by [*E.B.Burgina*] on the Fourier spectrometer 100/S BRUKER, as the excitation source was used the line of 1064 nm Nd-YAG-laser with power of 100 mW (fig.9). These spectra were discussed in detail in [15], in this case it should be noted only the growth of D – band for a $^{13}$C graphite composite and for glassy carbon. That is clear without saying in respect to the small-grained morphologically complex structure of these materials.

At the same time, in contrast to the glassy carbon, the spectrum of the $^{13}$C carbon composite demonstrates a visible overtone with a frequency of 2495 cm$^{-1}$ (fig.9). It is worth noting that the isotope shift of about 40 cm$^{-1}$ of the main bands for the $^{13}$C carbon composite is proportional to $\Delta\omega_{(q)} \sim \sqrt{13/12} = 1.041$.

### 7. The heat transport mechanism

As follows from the Hall effect measurements [15], the concentration of charge carriers in the $^{13}$C carbon composite does not exceed $4 \times 10^{19}$ cm$^{-3}$. It is almost the same or even lower than that one for the fine-grained MPG-6 graphite which is by four orders of magnitude smaller than the charge carriers concentration in e.g., copper. In our case, the Wiedemann-Franz ratio exceeds almost hundred times that of conventional metals. It is assumed then that the phonon contribution is dominant over the temperature range of heat transport for both the ordinary graphite and so as in the $^{13}$C composite.

The phonon thermal conductivity is of the form [43, 33]:

$$K_p = \Sigma_j \int C_j(\omega) \upsilon^2_j(\omega) \tau_j(\omega) d\omega. \qquad (5).$$

where *j* is the phonon polarization branch, i.e., two transverse acoustic branches and one longitudinal acoustic branch; $\upsilon_j$ is the phonon group velocity, which, in many solids, can be approximated by sound velocity; $\tau_j$ is the phonon relaxation time, $\omega$ is the phonon frequency and *C* is the heat capacity. The phonon mean, free path $\Lambda$ is related to the relaxation time as:

$$\Lambda = \tau \times \upsilon. \qquad (6)$$

In the relaxation-time approximation, the various scattering mechanisms, which limit Λ, are additive

$$\tau^{-1} = \Sigma \tau_i^{-1} \quad (7),$$

where $i$ enumerates the scattering processes.

In typical solids, the acoustic phonons, which carry the bulk of heat, are scattered by other phonons, lattice defects, impurities, conduction electrons, and interfaces.

In [41], the theoretical expressions were derived for the phonon thermal conductivity of single crystals of graphite in the basal plane at room and at elevated temperatures. The phonons were treated by a two-dimensional Debye model in the frequency range from 4 to 46 THz. The mean free paths are calculated for anharmonic three-phonon interactions, scattering by point defects and scattering by grain boundaries. The thermal conductivity is proportional to the basal plane (*a*-plane) and for a single crystal is larger, by a factor of the order of 100 than that in the *c*-axis direction. Thus, the thermal conductivity of polycrystalline aggregates depends on the nature of the structure [42]. More significant than the numerical value is the spectral distribution of the heat transport, for this determines how sensitive the thermal conductivity is to boundaries and defects. In particular, because much of the heat is carried by frequencies near lower frequency limit $\omega_c$, where the intrinsic mean free path (MFP) is relatively long – on the order of 1 μm at 300 K – the conductivity is sensitive to the limits imposed by the grain size. The room temperature thermal conductivity is reduced if the grain size in the *a*-plane is below 2 μm.

There is a folded (*LA*) branch in the *c*-direction and a transverse, optical (*TO*) branch in the *a*-direction. These modes should interact strongly with the *a*-direction acoustic modes. Therefore, no significant heat transport in the *a*-direction should be expected at either less than 4 THz or less than $\omega_c = 2.5 \times 10^{13}$ s$^{-1}$. It is assumed then [41] that there are two branches that make the major contribution to the heat current: the longitudinal branch and the fast transverse branch. The slow transversal branch has an unusual dispersion with a low average velocity and its contribution is neglected. In this highly simplified graphite model, the thermal conductivity is determined by the Debye velocity as the average sound speed of the *LA* and *TA* acoustic modes.

Thus, Klemens used the Debye velocity instead of the group velocity and the Grüneisen parameter γ has to be 2 (the experimental value of graphite). Moreover, using the specific heat value for high temperatures ($C \sim \omega$) and assuming the classical phonons distribution, he estimated the thermal conductivity of graphite, which is close to the experimental one for the highly oriented pyrolytic graphite (HOPG), as ~ 1900 W/m×K. Fig. 10 shows the thermal conductivity calculations by relaxation time approach of anharmonic three-phonon processes [43,33].

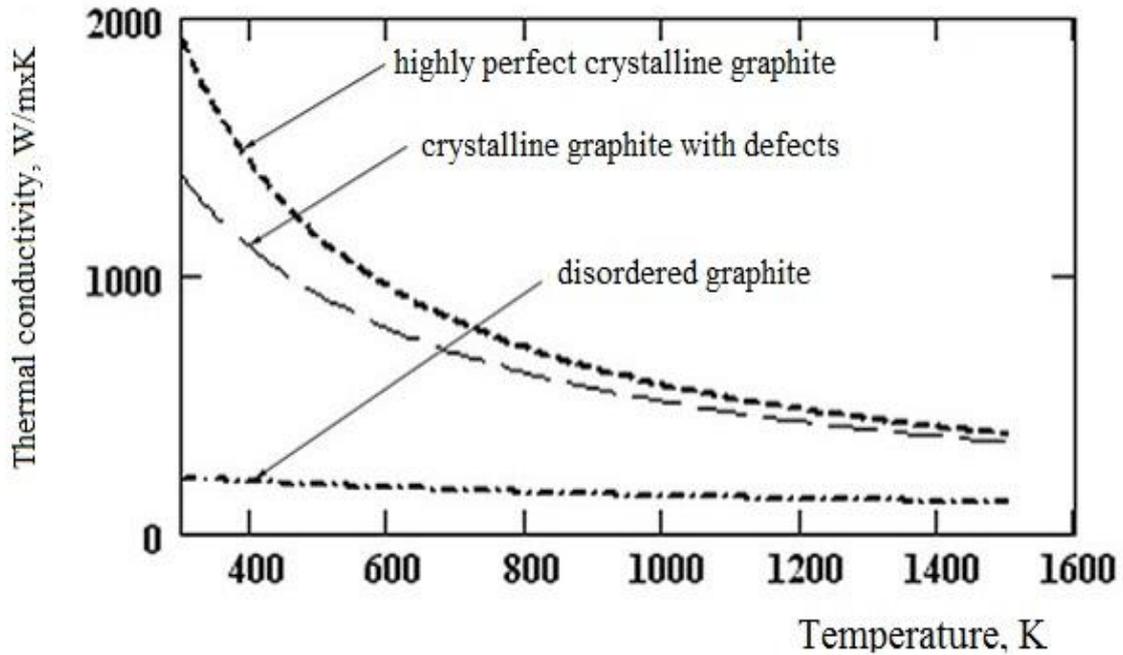

**Fig**. **10** *The thermal conductivity was calculated by relaxation time approach of anharmonic three-phonon processes model for crystal graphite (1). In this case, the defects (dislocation) and small grain size are taken into account. The calculations agree with* [41, 51].

A number of published papers [52, 53] currently study the influence of isotope effect on the thermal conductivity of graphite. For example, in [53], the molecular dynamic method was used to show that in the graphene layer the thermal conductivity can be reduced to 80% by random substitution of the $^{nat}C$ atoms by the $^{13}C$ isotope. The largest reduction would take a place if the total concentration of the substitution atoms of the isotope $^{13}C$ is approximately half the total number of carbon atoms in the graphene layer.

It should be noted, however, that the data on the thermal conductivity of traditional graphite, such as SGL or MPG-6, does not differ from the reference data of the manufacturer. There is no doubt that over the measured temperature range, a monotonic decrease in thermal conductivity with temperature is mainly due to both the three phonon–phonon interaction and the scattering of phonons at the boundary of the crystallite grain or the various kinds of defects.

8. **Thermal expansion measurements**

The experimental study of thermal expansion of the $^{13}C$ carbon composite was carried out in [54]. The data on the thermal expansion for various brands of graphite are shown in fig. 11.

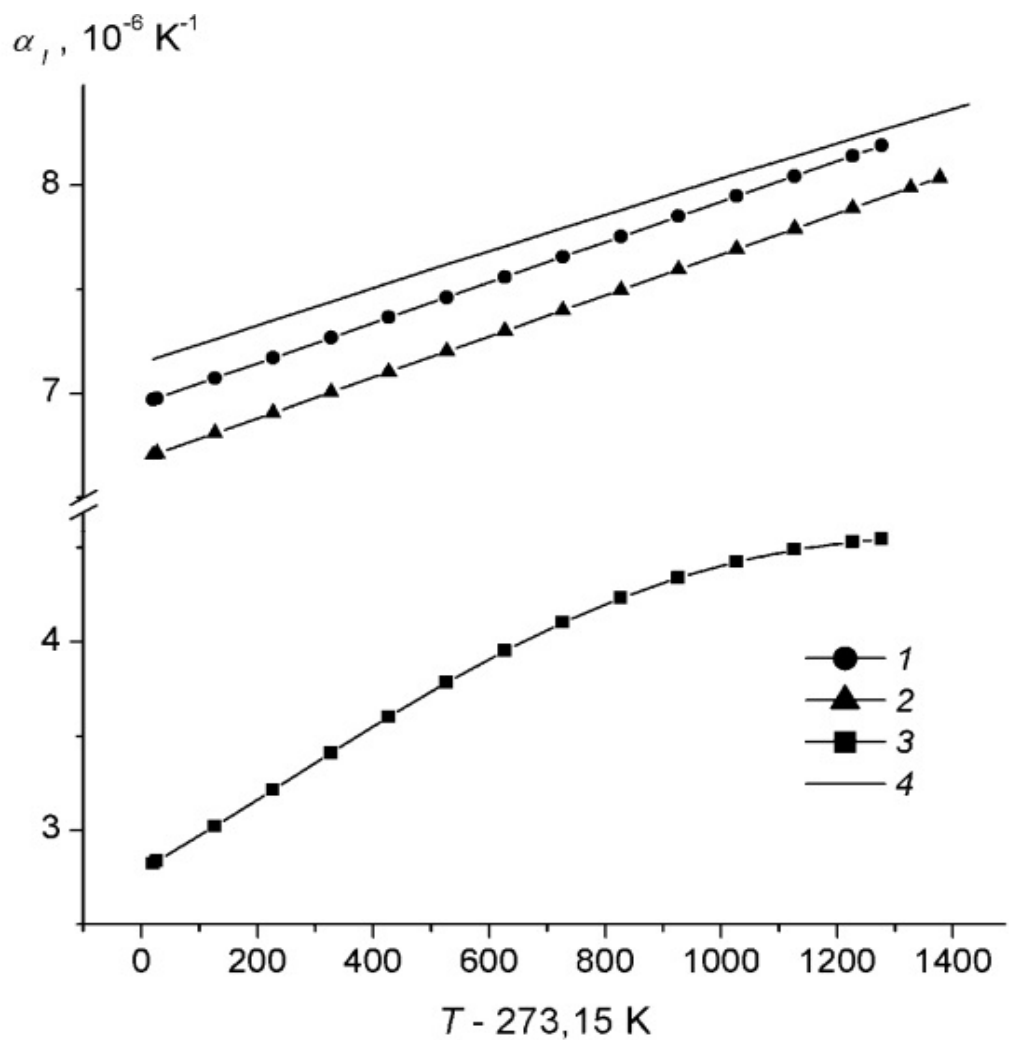

Fig. 11. *Average integral coefficient of linear expansion of graphite. 1) - POCO AXF-5Q, 2) - MPG-6, 3) - $^{13}C$ carbon composite, 4) - POCO AXM-5Q* [54].

It is clearly seen, that dense polycrystalline graffiti POCO and MPG-6 brands have a similar average integral coefficients of linear expansion and practically identical temperature coefficients of expansion. The difference between integral linear expansion coefficient (coefficient of linear expansion) of the reference data for POCO AXM-5Q [55] and *POCO AXF-5Q* does not exceed $1,9 \times 10^{-7}$ $K^{-1}$ and $4,6 \times 10^{-7}$ $K^{-1}$ respectively. The temperature dependence of the coefficient of linear expansion for $^{13}C$ carbon composite is essentially nonlinear and about 45-65% less in absolute value than the first ones.

Low thermal expansion coefficient $^{13}C$ is not surprising, if one take into account differences between of the graphite microstructure. The size of the coherent scattering region (CSR) in the $^{13}C$ composites according to HRTEM measurements is about of 100Å, while the electrophysical measurements are shown ~ 150Å [15]. Therefore, the average size of micro-crystallites of the $^{13}C$ composite can be estimated of 10-15 nm value.

Open porosity of MPG-6 polycrystalline graphite, that was measured by mercury porosimetry method, is about $9 \times 10^{-2}$ cm$^3$/g. The average pore radius is of 1 micron about [56]. The open porosity of the more friable $^{13}$C composite is almost four hundred times exceeds the total pore volume of the MPG-6 composite. This way, the maximum of the micro-pores distribution of the $^{13}$C composite have a diameter about of 0.3-0.5 microns. Furthermore, another two peaks of the pore distribution are in the $^{13}$C composite: the micropores with a diameter of about 2 nm and mesopores with an average diameter of about 10 nm.

The coefficient of a thermal expansion of the carbon composite is determined by two competing factors: the thermal expansion of micro-crystallites and the presence of the pores, some long cracks and other irregularities of structure that can compensate for this expansion [57]. The coefficients of linear and volumetric expansion of turbostratic graphite will be always lower than those ones of the perfect monocrystals or polycrystals because of the its high porosity [40]. Furthermore, if the carbon material contains an in-ordered amorphous phase, that located between of crystallites, it also leads to a decrease of the thermal expansion coefficient.

## Conclusions

The X-ray diffraction and transmission electron microscopy measurements have shown that $^{13}$C carbon composites samples have a clearly pronounced turbostratic structure. In general, the internal structure of the $^{13}$C composites is quite complicated and consists of a several noticeably differing morphological forms of carbon. The roentgen fluorescence CK$\alpha$-spectra of the pure $^{13}$C powder and $^{13}$C composites are considerably distinguished from the spectrum of natural graphite. Quantum chemical calculations of C$_{150}$ graphene was shown that the increase in the density of $\sigma$-states is provided by the $\sigma$-electrons of broken bonds of the boundary atoms of carbon particle with the size of about 20 Å.

It was also shown that the temperature dependence of the $^{13}$C composite conductivity is determined by the quantum mechanical effect of the two-dimensional weak localization of charge carriers. Data of the thermal conductivity are compared with previous X-ray diffraction data and with the results of high-resolution electron microscopy. It was shown also that in all cases the heat transfer in graphite are caused by the phonons.

## Acknowledgement

Author is grateful to prof. L.B. Zuev and prof. Y.P. Sharkeev (ISPMS SB RAS) for unchanging attention and interest to the study; prof. S.V. Tsybulya with co-workers for X-Ray and HRTEM measurements; prof. A.I. Romanenko with co-workers for electrophysical measurements; prof. S.V. Stankus with co-workers for thermophysical measurements. Author is grateful as well as of laboratory assistants, technicians and engineers of BINP SB RAS, all others who were helped in the measurement and testing of the graphite samples.

## Refernces